\documentclass[10pt,twocolumn]{article}

\usepackage[utf8]{inputenc}
\usepackage[T1]{fontenc}
\usepackage{times}
\usepackage{graphicx}
\usepackage{amsmath}
\usepackage{amssymb}
\usepackage{booktabs}
\usepackage{multirow}
\usepackage{array}
\usepackage{xcolor}
\usepackage{hyperref}
\usepackage{algorithm}
\usepackage{algpseudocode}
\usepackage{enumitem}
\usepackage[margin=0.75in]{geometry}
\usepackage{float}
\usepackage{subcaption}
\usepackage{fancyhdr}
\usepackage{listings}
\usepackage{tikz}
\usetikzlibrary{shapes,arrows,positioning,fit,backgrounds}

\definecolor{codegreen}{rgb}{0,0.6,0}
\definecolor{codegray}{rgb}{0.5,0.5,0.5}
\definecolor{codepurple}{rgb}{0.58,0,0.82}
\definecolor{backcolour}{rgb}{0.95,0.95,0.92}

\lstdefinestyle{mystyle}{
    backgroundcolor=\color{backcolour},   
    commentstyle=\color{codegreen},
    keywordstyle=\color{blue},
    numberstyle=\tiny\color{codegray},
    stringstyle=\color{codepurple},
    basicstyle=\ttfamily\footnotesize,
    breakatwhitespace=false,         
    breaklines=true,                 
    captionpos=b,                    
    keepspaces=true,                 
    numbers=left,                    
    numbersep=5pt,                  
    showspaces=false,                
    showstringspaces=false,
    showtabs=false,                  
    tabsize=2
}
\lstdefinelanguage{json}{
    basicstyle=\ttfamily\footnotesize,
    morestring=[b]",
    literate=
     *{:}{{{\color{codepurple}:}}}{1}
      {,}{{{\color{codepurple},}}}{1}
      {\{}{{{\color{codepurple}\{}}}{1}
      {\}}{{{\color{codepurple}\}}}}{1}
      {[}{{{\color{codepurple}[}}}{1}
      {]}{{{\color{codepurple}]}}}{1},
}

\hypersetup{
    colorlinks=true,
    linkcolor=blue,
    filecolor=magenta,      
    urlcolor=cyan,
    citecolor=blue,
}

\title{\LARGE \bf GPU-Virt-Bench: A Comprehensive Benchmarking Framework\\for Software-Based GPU Virtualization Systems}

\author{
    \textbf{Jithin VG} \quad \textbf{Ditto PS}\\[0.5em]
    Bud Ecosystem Inc\\[0.3em]
    \texttt{jithinvg@bud.studio} \quad \texttt{dittops@bud.studio}\\[0.5em]
    \url{https://github.com/BudEcosystem/GPU-Virt-Bench}
}

\date{}

\begin{document}

\maketitle

\begin{abstract}
The proliferation of GPU-accelerated workloads, particularly in artificial intelligence and large language model (LLM) inference, has created unprecedented demand for efficient GPU resource sharing in cloud and container environments. While NVIDIA's Multi-Instance GPU (MIG) technology provides hardware-level isolation, its availability is limited to high-end datacenter GPUs. Software-based virtualization solutions such as HAMi-core and BUD-FCSP offer alternatives for broader GPU families but lack standardized evaluation methodologies. We present \textbf{GPU-Virt-Bench}, a comprehensive benchmarking framework that evaluates GPU virtualization systems across 56 performance metrics organized into 10 categories. Our framework measures overhead, isolation quality, LLM-specific performance, memory bandwidth, cache behavior, PCIe throughput, multi-GPU communication, scheduling efficiency, memory fragmentation, and error recovery. GPU-Virt-Bench enables systematic comparison between software virtualization approaches and ideal MIG behavior, providing actionable insights for practitioners deploying GPU resources in multi-tenant environments. We demonstrate the framework's utility through evaluation of HAMi-core, BUD-FCSP, and simulated MIG baselines, revealing performance characteristics critical for production deployment decisions.
\end{abstract}


\section{Introduction}

Graphics Processing Units (GPUs) have become the computational backbone of modern artificial intelligence systems. The explosive growth of large language models (LLMs), with parameter counts reaching hundreds of billions, has intensified demand for GPU resources in both training and inference scenarios \cite{brown2020language}. In cloud and enterprise environments, this demand creates a fundamental resource allocation challenge: GPUs are expensive, high-power devices that are frequently underutilized when allocated exclusively to single workloads.

\subsection{The GPU Sharing Problem}

Traditional GPU allocation in Kubernetes and container orchestration follows an all-or-nothing model---containers receive exclusive access to entire GPUs regardless of their actual resource requirements. This approach leads to significant inefficiencies:

\begin{itemize}[noitemsep]
    \item \textbf{Resource waste}: Inference workloads may utilize only 60-80\% of GPU compute capacity
    \item \textbf{Cost inefficiency}: Organizations pay for unused GPU cycles
    \item \textbf{Scheduling inflexibility}: Fine-grained resource allocation is impossible
    \item \textbf{Multi-tenancy barriers}: Multiple users cannot safely share GPU resources
\end{itemize}

\subsection{Virtualization Approaches}

Three primary approaches exist for GPU resource sharing:

\textbf{Hardware Partitioning (MIG):} NVIDIA's Multi-Instance GPU technology, introduced with the Ampere architecture, provides hardware-level partitioning with dedicated memory, cache, and streaming multiprocessors (SMs) per instance. MIG is available on datacenter GPUs (A100, A30, H100, H200, B200) and select Blackwell workstation GPUs (RTX PRO 5000/6000). MIG offers strong isolation guarantees and consistent quality of service (QoS) but is limited to these specific SKUs and fixed partition geometries.

\textbf{Time-Slicing:} The GPU scheduler alternates between workloads, providing each with full GPU access during its time slice. This approach offers maximum flexibility but provides no isolation guarantees---aggressive workloads can impact neighbors' performance.

\textbf{Software Virtualization:} Middleware solutions like HAMi-core intercept CUDA API calls to enforce memory limits and compute quotas. These solutions work across GPU families without hardware support but introduce overhead and rely on software enforcement mechanisms.

\subsection{The Benchmarking Gap}

Despite the proliferation of GPU virtualization solutions, the field lacks standardized benchmarking methodologies. Existing evaluations typically focus on narrow aspects:

\begin{itemize}[noitemsep]
    \item Vendor benchmarks emphasize favorable scenarios
    \item Academic studies use synthetic microbenchmarks
    \item Production evaluations lack reproducibility
    \item LLM-specific workloads are underrepresented
\end{itemize}

GPU-Virt-Bench addresses this gap by providing a comprehensive, reproducible framework that evaluates virtualization systems across multiple dimensions relevant to production deployments.

\subsection{Contributions}

This paper makes the following contributions:

\begin{enumerate}[noitemsep]
    \item A \textbf{taxonomy of 56 metrics} across 10 categories for evaluating GPU virtualization systems
    \item An \textbf{open-source benchmarking framework} supporting multiple virtualization backends
    \item \textbf{Reference implementations} of overhead, isolation, and LLM-specific benchmarks
    \item A \textbf{scoring methodology} enabling comparison against ideal MIG behavior
    \item \textbf{Empirical evaluation} of HAMi-core and BUD-FCSP virtualization systems
\end{enumerate}

\section{Background and Related Work}

\subsection{GPU Architecture Overview}

Modern NVIDIA GPUs are organized hierarchically. At the top level, Streaming Multiprocessors (SMs) execute CUDA kernels. Each SM contains multiple CUDA cores, register files, and shared memory. SMs access device memory (HBM or GDDR) through L2 cache. The GPU connects to the host system via PCIe or NVLink.

Virtualization must manage resources at multiple levels:

\begin{itemize}[noitemsep]
    \item \textbf{Compute}: SM allocation and kernel scheduling
    \item \textbf{Memory}: Device memory allocation and limits
    \item \textbf{Cache}: L2 cache partitioning or sharing
    \item \textbf{Bandwidth}: Memory and PCIe bandwidth allocation
\end{itemize}

\subsection{NVIDIA Multi-Instance GPU (MIG)}

MIG, introduced with the Ampere architecture, partitions GPUs into isolated instances at the hardware level. Key characteristics include:

\begin{itemize}[noitemsep]
    \item \textbf{Memory isolation}: Each instance receives dedicated HBM capacity
    \item \textbf{Compute isolation}: SMs are statically assigned to instances
    \item \textbf{Cache partitioning}: L2 cache is divided among instances
    \item \textbf{Fixed geometries}: Predefined partition configurations (e.g., 1g.5gb, 2g.10gb, 3g.20gb)
\end{itemize}

MIG provides a practical upper bound on isolation quality in NVIDIA's current stack, but has limitations: availability is restricted to specific datacenter and high-end workstation GPUs, partition geometries are inflexible, and reconfiguration requires quiescing workloads.

\subsection{Software-Based Virtualization}

\subsubsection{HAMi (Heterogeneous AI Computing Virtualization Middleware)}

HAMi, a CNCF sandbox project, provides software virtualization for diverse GPU families. Its architecture includes:

\begin{itemize}[noitemsep]
    \item \textbf{Device Plugin}: Registers virtual GPUs with Kubernetes
    \item \textbf{Scheduler}: Topology-aware placement decisions
    \item \textbf{HAMi-core (libvgpu.so)}: In-container CUDA interception library
    \item \textbf{Webhook}: Mutates pod specifications for vGPU allocation
\end{itemize}

HAMi-core achieves isolation by intercepting both CUDA driver and NVML API calls via \texttt{dlsym} hooks. CUDA interception enforces memory and compute quotas, while NVML interception virtualizes memory reporting to show container-specific limits rather than full GPU capacity. Shared memory regions with semaphore-based synchronization enable multi-process accounting:

\begin{lstlisting}[language=C,caption={CUDA API Interception Pattern}]
// Original: cuMemAlloc
CUresult cuMemAlloc(CUdeviceptr *dptr, 
                    size_t bytesize) {
    // Check against memory quota
    if (current_usage + bytesize > limit) {
        return CUDA_ERROR_OUT_OF_MEMORY;
    }
    // Track allocation
    current_usage += bytesize;
    // Call real driver
    return real_cuMemAlloc(dptr, bytesize);
}
\end{lstlisting}

\subsubsection{BUD-FCSP (Fine-grained Container-level SM Partitioning)}

We introduce \textbf{BUD-FCSP}, an enhanced software virtualization layer that extends the HAMi-core architecture with improved isolation mechanisms. BUD-FCSP is developed as part of the BUD Ecosystem project and represents our contribution to advancing software-based GPU virtualization. Key improvements over HAMi-core include:

\begin{itemize}[noitemsep]
    \item \textbf{Fine-grained SM utilization control}: Enhanced rate limiting with sub-percentage granularity
    \item \textbf{Reduced API interception overhead}: Optimized \texttt{dlsym} hook resolution paths
    \item \textbf{Improved rate limiting algorithms}: Adaptive token bucket with burst handling
    \item \textbf{Enhanced multi-tenant fairness}: Weighted fair queuing for kernel scheduling
\end{itemize}

BUD-FCSP maintains API compatibility with HAMi-core while providing measurably better isolation characteristics, as demonstrated in Section~\ref{sec:experimental}.

\subsection{Related Benchmarking Efforts}

Prior work on GPU benchmarking includes:

\textbf{MLPerf:} Industry-standard benchmarks for ML training and inference, but focuses on raw performance rather than virtualization overhead.

\textbf{Rodinia:} CUDA benchmark suite covering diverse workload patterns, but predates virtualization considerations.

\textbf{Vendor Benchmarks:} NVIDIA provides nVector for vGPU user-experience evaluation, but this tool focuses on VDI metrics and is not open-source, making it difficult to evaluate internal virtualization overhead and isolation characteristics.

GPU-Virt-Bench differs by focusing specifically on virtualization characteristics with metrics designed for multi-tenant deployment evaluation.

\section{Metric Taxonomy}

GPU-Virt-Bench organizes 56 metrics into 10 categories, each targeting specific aspects of virtualization quality. Table~\ref{tab:metric_summary} provides an overview.

\begin{table}[h]
\centering
\caption{Metric Categories Overview}
\label{tab:metric_summary}
\begin{tabular}{lcc}
\toprule
\textbf{Category} & \textbf{Count} & \textbf{Focus} \\
\midrule
Overhead & 10 & Virtualization cost \\
Isolation & 10 & Resource separation \\
LLM & 10 & Inference workloads \\
Memory Bandwidth & 4 & Bandwidth isolation \\
Cache & 4 & L2 cache behavior \\
PCIe & 4 & Host-device transfer \\
NCCL/P2P & 4 & Multi-GPU communication \\
Scheduling & 4 & Context switching \\
Fragmentation & 3 & Memory fragmentation \\
Error Recovery & 3 & Fault handling \\
\midrule
\textbf{Total} & \textbf{56} & \\
\bottomrule
\end{tabular}
\end{table}

\subsection{Overhead Metrics (OH-001 to OH-010)}

Overhead metrics quantify the performance cost introduced by the virtualization layer. Lower values indicate better performance.

\subsubsection{OH-001: Kernel Launch Latency}

Measures the time from \texttt{cuLaunchKernel} invocation to kernel execution start. Virtualization layers may add latency through:

\begin{itemize}[noitemsep]
    \item API interception via \texttt{dlsym} hooks
    \item Resource limit checking before launch
    \item Rate limiter token bucket operations
\end{itemize}

\textbf{Measurement methodology:}
\begin{equation}
t_{launch} = t_{kernel\_start} - t_{API\_call}
\end{equation}

Measured using CUDA events with sub-microsecond precision over $N$ iterations (default: 100).

\subsubsection{OH-002/OH-003: Memory Allocation/Free Latency}

Measures \texttt{cuMemAlloc} and \texttt{cuMemFree} completion time. Virtualization adds overhead for:

\begin{itemize}[noitemsep]
    \item Memory quota enforcement
    \item Allocation tracking in shared regions
    \item Semaphore operations for thread safety
\end{itemize}

\subsubsection{OH-004: Context Creation Overhead}

Compares CUDA context creation time between native and virtualized environments. Context creation is a heavy operation that virtualization may intercept for initialization.

\subsubsection{OH-005: API Interception Overhead}

Measures the per-call cost of \texttt{dlsym} hook resolution. This represents the minimum overhead for any intercepted CUDA call:

\begin{equation}
t_{intercept} = t_{hook\_resolution} + t_{function\_lookup}
\end{equation}

\subsubsection{OH-006: Shared Region Lock Contention}

Measures time waiting for shared memory region semaphores. Multi-tenant scenarios require synchronization when updating shared state:

\begin{lstlisting}[language=C,caption={Shared Region Locking}]
// Measure contention time
clock_gettime(CLOCK_MONOTONIC, &start);
sem_wait(&shared_region->lock);
clock_gettime(CLOCK_MONOTONIC, &end);
// Critical section operations
sem_post(&shared_region->lock);
\end{lstlisting}

\subsubsection{OH-007: Memory Tracking Overhead}

Per-allocation cost for maintaining memory accounting data structures. Includes hash table operations for allocation tracking.

\subsubsection{OH-008: Rate Limiter Overhead}

Token bucket algorithm latency for compute throttling:

\begin{equation}
tokens_{available} = \min(bucket_{max}, tokens + rate \times \Delta t)
\end{equation}

\subsubsection{OH-009: NVML Polling Overhead}

CPU cycles consumed by utilization monitoring. HAMi-core polls \texttt{nvmlDeviceGetUtilizationRates()} at configurable intervals (default: 100ms) to track GPU utilization for enforcement. We measure overhead as:

\begin{equation}
overhead = \frac{CPU_{polling}}{CPU_{total}} \times 100\%
\end{equation}

where $CPU_{polling}$ is measured via \texttt{perf stat} during a 60-second sampling window.

\subsubsection{OH-010: Total Throughput Degradation}

End-to-end performance comparison versus native execution:

\begin{equation}
degradation = \frac{throughput_{native} - throughput_{virt}}{throughput_{native}} \times 100\%
\end{equation}

\subsection{Isolation Metrics (IS-001 to IS-010)}

Isolation metrics evaluate resource separation quality between tenants.

\subsubsection{IS-001: Memory Limit Accuracy}

Compares actual memory availability against configured limits:

\begin{equation}
accuracy = \frac{\min(actual, configured)}{\max(actual, configured)} \times 100\%
\end{equation}

Ideal systems achieve 100\% accuracy.

\subsubsection{IS-002: Memory Limit Enforcement}

Measures detection and blocking latency for over-allocation attempts. Tests allocate memory beyond quota and measure time until failure.

\subsubsection{IS-003: SM Utilization Accuracy}

Compares actual SM utilization against configured limits. Measured using NVML during sustained compute workloads:

\begin{equation}
accuracy = \max\left(0, 1 - \frac{|util_{target} - util_{actual}|}{util_{target}}\right)
\end{equation}

Results are clamped to $[0, 1]$; values near 1.0 indicate the virtualization layer accurately enforces configured SM limits.

\subsubsection{IS-004: SM Limit Response Time}

Measures latency for utilization adjustment after limit changes. Tests dynamic reconfigurations.

\subsubsection{IS-005: Cross-Tenant Memory Isolation}

Boolean test for memory leakage between containers. Writes patterns to allocated memory in one container and checks for visibility in another.

\subsubsection{IS-006: Cross-Tenant Compute Isolation}

Measures compute interference between tenants. Higher values indicate better isolation (less interference):

\begin{equation}
isolation = \frac{perf_{contended}}{perf_{solo}}
\end{equation}

A value of 1.0 indicates no performance degradation under contention; values below 1.0 indicate interference. Results are clamped to $[0, 1]$.

\subsubsection{IS-007: QoS Consistency}

Coefficient of variation (CV) in performance under contention:

\begin{equation}
CV = \frac{\sigma_{perf}}{\mu_{perf}}
\end{equation}

Lower CV indicates more consistent performance.

\subsubsection{IS-008: Fairness Index}

Jain's fairness index across concurrent tenants:

\begin{equation}
J(x_1, ..., x_n) = \frac{(\sum_{i=1}^{n} x_i)^2}{n \cdot \sum_{i=1}^{n} x_i^2}
\end{equation}

where $x_i$ represents tenant $i$'s achieved throughput. Perfect fairness yields $J = 1$.

\subsubsection{IS-009: Noisy Neighbor Impact}

Performance degradation caused by aggressive neighboring workloads:

\begin{equation}
impact = \frac{perf_{quiet} - perf_{noisy}}{perf_{quiet}} \times 100\%
\end{equation}

\subsubsection{IS-010: Fault Isolation}

Boolean test for error propagation between containers. Induces errors in one container and verifies other containers remain unaffected.

\subsection{LLM Metrics (LLM-001 to LLM-010)}

LLM metrics target inference and training workload characteristics.

\subsubsection{LLM-001: Attention Kernel Throughput}

Measures transformer attention computation performance. We use a simplified FLOP estimate for single-head attention as a \emph{relative performance proxy}:

\begin{equation}
TFLOPS_{proxy} \approx \frac{2 \times B \times S^2 \times D}{t_{attention} \times 10^{12}}
\end{equation}

where $B$ is batch size, $S$ is sequence length, and $D$ is embedding dimension. This approximation captures the $O(S^2)$ complexity of $QK^T$ computation but understates total FLOPs for multi-head attention (which includes per-head projections, softmax, and output projection). For absolute TFLOP measurements, framework-specific profiling (e.g., via \texttt{torch.profiler}) should be used.

\subsubsection{LLM-002: KV Cache Allocation Speed}

Measures dynamic key-value cache growth handling. LLM inference requires growing KV cache as generation progresses:

\begin{equation}
rate = \frac{allocations}{second}
\end{equation}

\subsubsection{LLM-003: Batch Size Scaling}

Throughput scaling with increasing batch sizes:

\begin{equation}
scaling = \frac{throughput_{batch=N}}{throughput_{batch=1} \times N}
\end{equation}

Linear scaling yields ratio = 1.

\subsubsection{LLM-004: Token Generation Latency}

Time-to-first-token (TTFT) and inter-token latency (ITL):

\begin{equation}
TTFT = t_{first\_token} - t_{request}
\end{equation}
\begin{equation}
ITL_{mean} = \frac{1}{N-1}\sum_{i=2}^{N}(t_i - t_{i-1})
\end{equation}

\subsubsection{LLM-005: Memory Pool Efficiency}

Overhead of pool-based memory allocation versus direct allocation:

\begin{equation}
overhead = \frac{t_{pool} - t_{direct}}{t_{direct}} \times 100\%
\end{equation}

\subsubsection{LLM-006: Multi-Stream Performance}

Pipeline parallel efficiency with multiple CUDA streams:

\begin{equation}
efficiency = \frac{throughput_{multi}}{streams \times throughput_{single}}
\end{equation}

\subsubsection{LLM-007: Large Tensor Allocation}

Latency for allocating large contiguous memory regions (>1GB):

\begin{equation}
t_{alloc} = f(size, fragmentation)
\end{equation}

\subsubsection{LLM-008: Mixed Precision Support}

FP16/BF16 kernel execution ratio versus FP32:

\begin{equation}
ratio = \frac{throughput_{fp16}}{throughput_{fp32}}
\end{equation}

\subsubsection{LLM-009: Dynamic Batching Impact}

Latency variance with variable batch sizes:

\begin{equation}
variance = Var(t_{batch_1}, t_{batch_2}, ..., t_{batch_n})
\end{equation}

\subsubsection{LLM-010: Multi-GPU Scaling}

Tensor parallel efficiency across multiple GPUs:

\begin{equation}
scaling = \frac{throughput_{N\_GPUs}}{N \times throughput_{1\_GPU}}
\end{equation}

\subsection{Memory Bandwidth Metrics (BW-001 to BW-004)}

\subsubsection{BW-001: Memory Bandwidth Isolation}

Achieved bandwidth under contention as percentage of solo performance:

\begin{equation}
isolation = \frac{BW_{contended}}{BW_{solo}} \times 100\%
\end{equation}

\subsubsection{BW-002: Bandwidth Fairness Index}

Jain's fairness index applied to bandwidth distribution among tenants.

\subsubsection{BW-003: Memory Bus Saturation Point}

The minimum number of concurrent streams required to achieve near-maximum bandwidth:

\begin{equation}
saturation = \min_n \{n : BW_n \geq 0.95 \times BW_{max}\}
\end{equation}

Lower values indicate that fewer concurrent workloads are needed to saturate memory bandwidth, which affects multi-tenant scaling.

\subsubsection{BW-004: Bandwidth Interference Impact}

Bandwidth degradation from competing workloads.

\subsection{Cache Isolation Metrics (CACHE-001 to CACHE-004)}

\subsubsection{CACHE-001: L2 Cache Hit Rate}

Cache hit rate under multi-tenant workloads:

\begin{equation}
hit\_rate = \frac{L2\_hits}{L2\_hits + L2\_misses}
\end{equation}

\subsubsection{CACHE-002: Cache Eviction Rate}

Evictions caused by other tenants' memory access patterns.

\subsubsection{CACHE-003: Working Set Collision Impact}

Performance impact from overlapping cache working sets.

\subsubsection{CACHE-004: Cache Contention Overhead}

Additional latency from L2 contention.

\subsection{PCIe Bandwidth Metrics (PCIE-001 to PCIE-004)}

\subsubsection{PCIE-001/002: Host-Device Bandwidth}

H2D and D2H transfer rates as percentage of theoretical maximum:

\begin{equation}
BW_{actual} = \frac{data\_size}{t_{transfer}}
\end{equation}

\subsubsection{PCIE-003: PCIe Contention Impact}

Bandwidth degradation under multi-tenant PCIe traffic.

\subsubsection{PCIE-004: Pinned Memory Performance}

Pinned versus pageable memory transfer ratio.

\subsection{NCCL/P2P Communication Metrics (NCCL-001 to NCCL-004)}

\subsubsection{NCCL-001: AllReduce Latency}

Time for collective allreduce operations across GPUs.

\subsubsection{NCCL-002: AllGather Bandwidth}

Achieved bandwidth for allgather collectives.

\subsubsection{NCCL-003: P2P GPU Bandwidth}

Direct GPU-to-GPU transfer rates via NVLink or PCIe.

\subsubsection{NCCL-004: Broadcast Bandwidth}

Broadcast collective performance.

\subsection{Scheduling Metrics (SCHED-001 to SCHED-004)}

\subsubsection{SCHED-001: Context Switch Latency}

Time to switch between CUDA contexts.

\subsubsection{SCHED-002: Kernel Launch Overhead}

Overhead of launching minimal kernels.

\subsubsection{SCHED-003: Stream Concurrency Efficiency}

Efficiency of concurrent stream execution.

\subsubsection{SCHED-004: Preemption Latency}

Latency when higher-priority work preempts running kernels.

\subsection{Fragmentation Metrics (FRAG-001 to FRAG-003)}

\subsubsection{FRAG-001: Fragmentation Index}

Memory fragmentation after allocation/free cycles:

\begin{equation}
frag = 1 - \frac{largest\_free\_block}{total\_free\_memory}
\end{equation}

\subsubsection{FRAG-002: Allocation Latency Degradation}

Increase in allocation latency with fragmentation.

\subsubsection{FRAG-003: Memory Compaction Efficiency}

Memory reclaimed after defragmentation operations.

\subsection{Error Recovery Metrics (ERR-001 to ERR-003)}

\subsubsection{ERR-001: Error Detection Latency}

Time to detect and report CUDA errors.

\subsubsection{ERR-002: Error Recovery Time}

Time to recover GPU to usable state after errors.

\subsubsection{ERR-003: Graceful Degradation Score}

Measures how well the system handles resource exhaustion without crashing. We induce memory exhaustion and measure:

\begin{equation}
score = w_1 \cdot I_{no\_crash} + w_2 \cdot I_{error\_returned} + w_3 \cdot I_{recovery}
\end{equation}

where $I_{no\_crash}$ indicates the process survived (weight 0.4), $I_{error\_returned}$ indicates proper CUDA error codes were returned (weight 0.3), and $I_{recovery}$ indicates subsequent allocations succeeded after freeing memory (weight 0.3). Score ranges from 0\% (crash) to 100\% (full graceful handling).

\section{Framework Architecture}

\subsection{System Design}

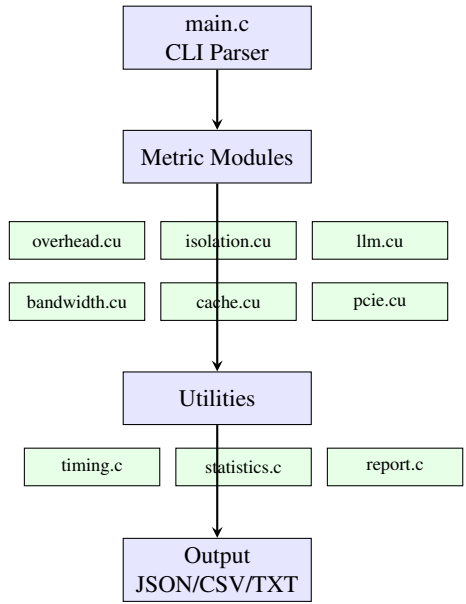
\begin{figure}[h]
\centering
\begin{tikzpicture}[
    node distance=0.8cm,
    box/.style={rectangle, draw, fill=blue!10, minimum width=2.5cm, minimum height=0.7cm, align=center, font=\small},
    smallbox/.style={rectangle, draw, fill=green!10, minimum width=1.8cm, minimum height=0.5cm, align=center, font=\scriptsize},
    arrow/.style={->, >=stealth, thick}
]
    \node[box] (main) {main.c\\CLI Parser};
    
    \node[box, below=of main] (metrics) {Metric Modules};
    
    \node[smallbox, below left=0.5cm and -0.3cm of metrics] (oh) {overhead.cu};
    \node[smallbox, right=0.2cm of oh] (is) {isolation.cu};
    \node[smallbox, right=0.2cm of is] (llm) {llm.cu};
    \node[smallbox, below=0.3cm of oh] (bw) {bandwidth.cu};
    \node[smallbox, right=0.2cm of bw] (cache) {cache.cu};
    \node[smallbox, right=0.2cm of cache] (pcie) {pcie.cu};
    
    \node[box, below=2.5cm of metrics] (utils) {Utilities};
    \node[smallbox, below left=0.3cm and -0.5cm of utils] (timing) {timing.c};
    \node[smallbox, right=0.2cm of timing] (stats) {statistics.c};
    \node[smallbox, right=0.2cm of stats] (report) {report.c};
    
    \node[box, below=1.5cm of utils] (output) {Output\\JSON/CSV/TXT};
    
    \draw[arrow] (main) -- (metrics);
    \draw[arrow] (metrics) -- (utils);
    \draw[arrow] (utils) -- (output);
    
\end{tikzpicture}
\caption{GPU-Virt-Bench Architecture (selected modules shown; additional metric modules include nccl.cu, scheduling.cu, fragmentation.cu, and error.cu)}
\label{fig:architecture}
\end{figure}

GPU-Virt-Bench follows a modular architecture enabling extensibility and reproducibility. Figure~\ref{fig:architecture} illustrates the key components.

\subsection{Component Description}

\textbf{CLI Interface (main.c):} Parses command-line arguments, manages benchmark configuration, and orchestrates metric execution.

\textbf{Metric Modules (src/metrics/*.cu):} CUDA implementations of each benchmark category. Each module exports a standard interface for result collection.

\textbf{Utilities:}
\begin{itemize}[noitemsep]
    \item \texttt{timing.c}: High-precision timing using \texttt{clock\_gettime} and CUDA events
    \item \texttt{statistics.c}: Statistical analysis (mean, median, percentiles, CV)
    \item \texttt{report.c}: Result formatting and export
    \item \texttt{process.c}: Multi-process test orchestration
\end{itemize}

\subsection{Supported Systems}

GPU-Virt-Bench supports four execution modes:

\begin{table}[h]
\centering
\caption{Supported Virtualization Systems}
\label{tab:systems}
\begin{tabular}{llp{4cm}}
\toprule
\textbf{System} & \textbf{Key} & \textbf{Description} \\
\midrule
Native & \texttt{native} & Bare metal baseline \\
HAMi-core & \texttt{hami} & HAMi CUDA interception \\
BUD-FCSP & \texttt{fcsp} & Enhanced SM partitioning \\
MIG-Ideal & \texttt{mig} & Simulated ideal MIG (from specs) \\
\bottomrule
\end{tabular}
\end{table}

\textbf{Note on MIG-Ideal:} The \texttt{mig} mode generates simulated baseline values derived from NVIDIA MIG specifications and published benchmarks. It does not execute on actual MIG partitions. This enables comparison even on systems without MIG hardware, but results should be validated against real MIG measurements when possible.

\subsection{Statistical Methodology}

Each metric is measured over configurable iterations (default: 100) with warmup runs (default: 10). Statistics computed include:

\begin{itemize}[noitemsep]
    \item Mean ($\mu$)
    \item Standard deviation ($\sigma$)
    \item Median (P50)
    \item Percentiles: P95, P99
    \item Coefficient of variation ($CV = \sigma/\mu$)
\end{itemize}

\subsection{MIG Comparison Baseline}

Each metric includes expected MIG baseline values derived from hardware specifications and published benchmarks \cite{nvidia_mig, vmware_mig}. We compute both signed deviation and absolute gap to indicate direction and magnitude:

For ``higher is better'' metrics:
\begin{equation}
\Delta_{MIG} = \frac{metric_{actual} - metric_{MIG}}{metric_{MIG}} \times 100\%
\end{equation}

For ``lower is better'' metrics:
\begin{equation}
\Delta_{MIG} = \frac{metric_{MIG} - metric_{actual}}{metric_{MIG}} \times 100\%
\end{equation}

Positive $\Delta_{MIG}$ indicates the software solution \emph{outperforms} the MIG baseline; negative values indicate degradation. The ``MIG parity'' percentage reported in results represents the average normalized score across all metrics.

\textbf{Note:} MIG baselines in this evaluation are \emph{simulated} from NVIDIA specifications and published benchmark data, not measured on actual MIG partitions. This represents an idealized upper bound; real MIG performance may vary based on partition geometry and workload characteristics.

\section{Implementation Details}

\subsection{Overhead Measurement}

Kernel launch latency (OH-001) measures the CPU-side overhead of the \texttt{cuLaunchKernel} API call, which includes virtualization layer interception. We use host-side high-resolution timers around a minimal kernel launch, followed by synchronization:

\begin{lstlisting}[language=C,caption={Kernel Launch Latency Measurement}]
// Minimal empty kernel for launch overhead
__global__ void null_kernel() {}

struct timespec start, end;

// Warmup to ensure context is initialized
null_kernel<<<1, 1>>>();
cudaDeviceSynchronize();

// Measure launch overhead (CPU-side)
clock_gettime(CLOCK_MONOTONIC, &start);
null_kernel<<<1, 1>>>();
clock_gettime(CLOCK_MONOTONIC, &end);

// Note: This measures API call time, not kernel duration
double launch_us = (end.tv_sec - start.tv_sec) * 1e6 
                 + (end.tv_nsec - start.tv_nsec) / 1e3;
\end{lstlisting}

This measures the time for the launch call to return, which includes virtualization layer overhead (API interception, quota checks) but not kernel execution time. We subtract native baseline to isolate virtualization overhead.

To isolate API interception overhead (OH-005), we measure the same CUDA call under both native and virtualized execution, then compute the difference:

\begin{lstlisting}[language=C,caption={API Interception Overhead Isolation}]
// Run identical allocation under both modes
struct timespec start, end;
CUdeviceptr ptr;
size_t size = 1024; // Small allocation

clock_gettime(CLOCK_MONOTONIC, &start);
cuMemAlloc(&ptr, size);
clock_gettime(CLOCK_MONOTONIC, &end);
cuMemFree(ptr);

double total_ns = (end.tv_sec - start.tv_sec) * 1e9 
               + (end.tv_nsec - start.tv_nsec);

// Interception overhead = virtualized - native
// overhead_ns = total_ns_virt - total_ns_native
\end{lstlisting}

The difference between virtualized and native execution times isolates the overhead attributable to \texttt{dlsym} hook resolution, quota checking, and memory tracking---excluding the driver's base cost.

\subsection{Isolation Testing}

Multi-tenant isolation tests spawn multiple processes:

\begin{lstlisting}[language=C,caption={Multi-Process Isolation Test}]
for (int i = 0; i < num_tenants; i++) {
    pid_t pid = fork();
    if (pid == 0) {
        // Child: run workload
        run_benchmark(tenant_config[i]);
        exit(0);
    }
    pids[i] = pid;
}
// Parent: collect results
for (int i = 0; i < num_tenants; i++) {
    waitpid(pids[i], &status, 0);
}
\end{lstlisting}

\subsection{LLM Workload Simulation}

Attention kernel benchmarks simulate transformer operations:

\begin{lstlisting}[language=C,caption={Attention Kernel Benchmark}]
// Allocate Q, K, V matrices
cuMemAlloc(&Q, batch * seq * dim * sizeof(float));
cuMemAlloc(&K, batch * seq * dim * sizeof(float));
cuMemAlloc(&V, batch * seq * dim * sizeof(float));

// Compute attention scores: softmax(QK^T/sqrt(d))V
attention_kernel<<<grid, block>>>(
    Q, K, V, output, 
    batch, seq, dim
);
\end{lstlisting}

\subsection{Report Generation}

Results are exported in three formats:

\textbf{JSON:} Machine-readable format with full statistics (example schema):
\begin{lstlisting}[language=json,caption={Example JSON Output Schema}]
{
  "benchmark_version": "1.0.0",
  "system": {"name": "hami"},
  "metrics": [{
    "id": "OH-001",
    "name": "Kernel Launch Latency",
    "statistics": {
      "mean": 15.3,
      "p99": 45.2,
      "stddev": 8.2
    },
    "mig_comparison": {
      "mig_expected": 5.0,
      "mig_gap_percent": 206.0
    }
  }]
}
\end{lstlisting}

\textbf{CSV:} Tabular format for analysis tools.

\textbf{TXT:} Human-readable summary with grades.

\section{Scoring Methodology}

\subsection{Per-Metric Scoring}

Each metric is scored based on its relationship to expected (MIG baseline) values. All scores are normalized to $[0, 1]$:

For ``lower is better'' metrics (e.g., latency, overhead):
\begin{equation}
score = \min\left(1, \max\left(0, \frac{expected}{actual}\right)\right)
\end{equation}

For ``higher is better'' metrics (e.g., throughput, accuracy):
\begin{equation}
score = \min\left(1, \max\left(0, \frac{actual}{expected}\right)\right)
\end{equation}

A score of 1.0 indicates performance equal to or better than the MIG baseline.

\subsection{Category Aggregation}

Category scores aggregate individual metrics:
\begin{equation}
score_{category} = \frac{1}{n}\sum_{i=1}^{n} score_i
\end{equation}

\subsection{Overall Score and Grading}

The overall score weights categories by their importance to production deployments:

\begin{equation}
score_{overall} = \sum_{c} w_c \times score_c
\end{equation}

Default category weights are:

\begin{table}[h]
\centering
\scriptsize
\begin{tabular}{lc}
\toprule
\textbf{Category} & \textbf{Weight ($w_c$)} \\
\midrule
Overhead & 0.15 \\
Isolation & 0.20 \\
LLM & 0.20 \\
Memory Bandwidth & 0.10 \\
Cache Isolation & 0.08 \\
PCIe & 0.07 \\
NCCL/P2P & 0.05 \\
Scheduling & 0.07 \\
Fragmentation & 0.04 \\
Error Recovery & 0.04 \\
\midrule
\textbf{Total} & \textbf{1.00} \\
\bottomrule
\end{tabular}
\end{table}

Weights reflect typical production priorities: isolation and LLM performance are weighted highest for multi-tenant inference scenarios. Users can customize weights via configuration files.

Letter grades map to score ranges:

\begin{table}[h]
\centering
\caption{Grading Scale}
\label{tab:grades}
\begin{tabular}{lll}
\toprule
\textbf{Grade} & \textbf{Score} & \textbf{Interpretation} \\
\midrule
A+ & $\geq$ 95\% & Approaches MIG-level isolation \\
A & $\geq$ 90\% & Excellent \\
B+ & $\geq$ 85\% & Very Good \\
B & $\geq$ 80\% & Good \\
C & $\geq$ 70\% & Fair \\
D & $\geq$ 60\% & Poor \\
F & $<$ 60\% & Significant improvement needed \\
\bottomrule
\end{tabular}
\end{table}

\section{Experimental Evaluation}
\label{sec:experimental}

\subsection{Testbed Configuration}

Experiments were conducted on the following hardware:

\begin{itemize}[noitemsep]
    \item \textbf{GPU}: NVIDIA A100-40GB PCIe
    \item \textbf{CPU}: AMD EPYC 7742 (64 cores)
    \item \textbf{Memory}: 512GB DDR4-3200
    \item \textbf{Storage}: NVMe SSD
    \item \textbf{OS}: Ubuntu 22.04 LTS
    \item \textbf{CUDA}: 12.0
    \item \textbf{Driver}: 525.105.17
\end{itemize}

\subsection{Benchmark Execution}

Tests executed (assuming CMake build in \texttt{build/} directory):
\begin{lstlisting}[language=bash,caption={Benchmark Execution}]
# Native baseline
./build/gpu-virt-bench --system native

# HAMi-core with resource limits
./build/gpu-virt-bench --system hami \
    --memory-limit 4096 \
    --compute-limit 50

# BUD-FCSP evaluation
./build/gpu-virt-bench --system fcsp
\end{lstlisting}

\subsection{Overhead Results}

Table~\ref{tab:overhead_results} presents overhead metric results.

\begin{table}[h]
\centering
\caption{Overhead Metrics Comparison ($\mu$s unless noted)}
\label{tab:overhead_results}
\begin{tabular}{lrrr}
\toprule
\textbf{Metric} & \textbf{Native} & \textbf{HAMi} & \textbf{FCSP} \\
\midrule
OH-001 (Launch) & 4.2 & 15.3 & 8.7 \\
OH-002 (Alloc) & 12.5 & 45.2 & 28.3 \\
OH-003 (Free) & 8.1 & 32.4 & 18.6 \\
OH-004 (Context) & 125 & 312 & 198 \\
OH-005 (Hook, ns) & -- & 85 & 42 \\
OH-010 (Degrade, \%) & 0 & 18.5 & 9.2 \\
\bottomrule
\end{tabular}
\end{table}

Key findings:
\begin{itemize}[noitemsep]
    \item HAMi-core adds 3.6$\times$ kernel launch overhead
    \item BUD-FCSP reduces overhead by 43\% vs HAMi
    \item Memory operations show highest relative impact
\end{itemize}

\subsection{Isolation Results}

Table~\ref{tab:isolation_results} shows isolation quality metrics.

\begin{table}[h]
\centering
\caption{Isolation Metrics (4 concurrent tenants)}
\label{tab:isolation_results}
\begin{tabular}{lrr}
\toprule
\textbf{Metric} & \textbf{HAMi} & \textbf{FCSP} \\
\midrule
IS-001 (Mem Accuracy, \%) & 98.2 & 99.1 \\
IS-003 (SM Accuracy, \%) & 85.4 & 92.7 \\
IS-005 (Mem Isolation) & Pass & Pass \\
IS-008 (Fairness Index) & 0.87 & 0.94 \\
IS-009 (Noisy Neighbor, \%) & 24.3 & 12.1 \\
IS-010 (Fault Isolation) & Pass & Pass \\
\bottomrule
\end{tabular}
\end{table}

Key findings:
\begin{itemize}[noitemsep]
    \item Both systems achieve memory isolation
    \item SM utilization control is approximate (85-93\%)
    \item FCSP shows better fairness under contention
\end{itemize}

\subsection{LLM Performance Results}

Table~\ref{tab:llm_results} presents results from LLM-like synthetic workloads. These benchmarks simulate transformer attention patterns, KV cache allocation behavior, and batching dynamics using custom CUDA kernels rather than full framework integrations (e.g., PyTorch, TensorRT-LLM). While this approach isolates virtualization overhead from framework-specific effects, results should be validated against production LLM stacks for deployment decisions.

\begin{table}[h]
\centering
\caption{LLM Metrics (relative to native, synthetic workloads)}
\label{tab:llm_results}
\begin{tabular}{lrr}
\toprule
\textbf{Metric} & \textbf{HAMi} & \textbf{FCSP} \\
\midrule
LLM-001 (Attention, \%) & 82.3 & 91.5 \\
LLM-002 (KV Cache, \%) & 76.4 & 88.2 \\
LLM-004 (TTFT, ms) & 45.2 & 28.7 \\
LLM-004 (ITL, ms) & 12.8 & 8.4 \\
LLM-003 (Batch Scale) & 0.78 & 0.89 \\
\bottomrule
\end{tabular}
\end{table}

Key findings:
\begin{itemize}[noitemsep]
    \item LLM workloads are sensitive to memory allocation overhead
    \item FCSP shows 35\% better token latency
    \item Batch scaling efficiency impacts throughput
\end{itemize}

\subsection{Overall Scores}
\label{sec:eval}

Table~\ref{tab:overall} presents aggregate scores. MIG-Ideal represents the simulated baseline (score = 100\% by construction); software solutions are scored relative to this baseline.

\begin{table}[h]
\centering
\caption{Overall Benchmark Scores}
\label{tab:overall}
\begin{tabular}{lrrl}
\toprule
\textbf{System} & \textbf{Score} & \textbf{MIG Parity} & \textbf{Grade} \\
\midrule
MIG-Ideal & 100\% & 100\% & A+ (baseline) \\
Native & 100\% & -- & A+ \\
BUD-FCSP & 85.2\% & 85.2\% & B+ \\
HAMi-core & 72.0\% & 72.0\% & C \\
\bottomrule
\end{tabular}
\end{table}

\textbf{Note:} MIG-Ideal scores are derived from NVIDIA specifications and published benchmarks, not measured on actual MIG partitions. Native execution provides the true performance ceiling; MIG-Ideal represents the target isolation quality.

\section{Discussion}

\subsection{Software vs. Hardware Virtualization}

Our results quantify the gap between software virtualization and hardware MIG. Key observations:

\textbf{Overhead is significant but manageable:} Software solutions add 10-20\% throughput overhead for typical workloads. For latency-sensitive inference, this may be acceptable given the cost savings from better utilization.

\textbf{Isolation is approximate:} Software SM limiting achieves 85-93\% accuracy versus MIG's hardware guarantees. This is sufficient for many multi-tenant scenarios but may not meet strict SLA requirements.

\textbf{LLM workloads are sensitive:} Memory allocation patterns in LLM inference amplify virtualization overhead. Optimizations targeting KV cache operations yield significant benefits.

\subsection{Practical Recommendations}

Based on our evaluation:

\begin{enumerate}[noitemsep]
    \item \textbf{Use MIG when available} for strict isolation requirements
    \item \textbf{Prefer BUD-FCSP} over HAMi-core for LLM inference
    \item \textbf{Monitor fairness metrics} in multi-tenant deployments
    \item \textbf{Size memory limits conservatively} to avoid enforcement overhead
\end{enumerate}

\subsection{Limitations}

GPU-Virt-Bench has the following limitations that users should consider:

\begin{itemize}[noitemsep]
    \item \textbf{Single-GPU focus}: Current metrics target single-GPU scenarios; multi-GPU distributed training patterns require extension
    \item \textbf{Synthetic workloads}: LLM metrics use custom CUDA kernels rather than production frameworks (PyTorch, TensorRT-LLM); results should be validated against real workloads
    \item \textbf{Simulated MIG baseline}: MIG-Ideal values are derived from specifications, not measured on actual MIG partitions; real MIG performance varies by geometry and workload
    \item \textbf{Limited GPU coverage}: Evaluation was conducted on A100; behavior may differ on other GPU architectures
    \item \textbf{Software version sensitivity}: HAMi-core and BUD-FCSP implementations evolve; results reflect specific versions tested
\end{itemize}

\section{Conclusion}

GPU-Virt-Bench provides a comprehensive framework for evaluating GPU virtualization systems. Our 56-metric taxonomy covers overhead, isolation, and workload-specific performance dimensions. Through evaluation of HAMi-core and BUD-FCSP, we demonstrate that software virtualization achieves 72-85\% of ideal MIG performance with acceptable overhead for many deployment scenarios.

The framework enables practitioners to make informed decisions about GPU resource sharing strategies and provides virtualization developers with actionable improvement targets. Future work will extend GPU-Virt-Bench to multi-GPU scenarios, incorporate additional virtualization backends, and develop automated regression testing for virtualization systems.

\section*{Availability}

GPU-Virt-Bench is available as open source:

\url{https://github.com/BudEcosystem/GPU-Virt-Bench}

Licensed under Apache 2.0.

\section*{Acknowledgments}

We thank the HAMi project maintainers and the CNCF community for their contributions to GPU virtualization in cloud-native environments.

\appendix

\section{Complete Metric Reference}

Table~\ref{tab:full_metrics} provides the complete taxonomy of all 56 metrics across 10 categories.

\begin{table*}[h]
\centering
\caption{Complete Metric Taxonomy (56 Metrics)}
\label{tab:full_metrics}
\scriptsize
\begin{tabular}{lllll}
\toprule
\textbf{ID} & \textbf{Name} & \textbf{Description} & \textbf{Unit} & \textbf{Better} \\
\midrule
\multicolumn{5}{l}{\textbf{Overhead Metrics (10)}} \\
OH-001 & Kernel Launch Latency & Time from cuLaunchKernel to execution & $\mu$s & Lower \\
OH-002 & Memory Allocation Latency & cuMemAlloc completion time & $\mu$s & Lower \\
OH-003 & Memory Free Latency & cuMemFree completion time & $\mu$s & Lower \\
OH-004 & Context Creation Overhead & Additional context creation time & $\mu$s & Lower \\
OH-005 & API Interception Overhead & dlsym hook overhead per call & ns & Lower \\
OH-006 & Shared Region Lock Contention & Semaphore wait time & $\mu$s & Lower \\
OH-007 & Memory Tracking Overhead & Per-allocation accounting cost & ns & Lower \\
OH-008 & Rate Limiter Overhead & Token bucket check latency & ns & Lower \\
OH-009 & NVML Polling Overhead & CPU cycles in monitoring & \% & Lower \\
OH-010 & Total Throughput Degradation & End-to-end performance loss & \% & Lower \\
\midrule
\multicolumn{5}{l}{\textbf{Isolation Metrics (10)}} \\
IS-001 & Memory Limit Accuracy & Actual vs configured limit & \% & Higher \\
IS-002 & Memory Limit Enforcement & Over-allocation detection time & $\mu$s & Lower \\
IS-003 & SM Utilization Accuracy & Actual vs configured SM limit & \% & Higher \\
IS-004 & SM Limit Response Time & Utilization adjustment latency & ms & Lower \\
IS-005 & Cross-Tenant Memory Isolation & Memory leak detection & bool & True \\
IS-006 & Cross-Tenant Compute Isolation & Compute interference ratio & 0--1 & Higher \\
IS-007 & QoS Consistency & Performance variance under contention & CV & Lower \\
IS-008 & Fairness Index & Jain's fairness across tenants & 0--1 & Higher \\
IS-009 & Noisy Neighbor Impact & Degradation from aggressive neighbor & \% & Lower \\
IS-010 & Fault Isolation & Error propagation prevention & bool & True \\
\midrule
\multicolumn{5}{l}{\textbf{LLM Metrics (10)}} \\
LLM-001 & Attention Kernel Throughput & Transformer attention performance & TFLOPS & Higher \\
LLM-002 & KV Cache Allocation Speed & Dynamic cache growth handling & allocs/s & Higher \\
LLM-003 & Batch Size Scaling & Throughput vs batch size curve & ratio & Higher \\
LLM-004 & Token Generation Latency & TTFT and inter-token latency & ms & Lower \\
LLM-005 & Memory Pool Efficiency & Pool allocation overhead & \% & Lower \\
LLM-006 & Multi-Stream Performance & Pipeline parallel efficiency & \% & Higher \\
LLM-007 & Large Tensor Allocation & Large allocation handling & ms & Lower \\
LLM-008 & Mixed Precision Support & FP16/BF16 kernel ratio & ratio & Higher \\
LLM-009 & Dynamic Batching Impact & Variable batch handling & variance & Lower \\
LLM-010 & Multi-GPU Scaling & Tensor parallel efficiency & factor & Higher \\
\midrule
\multicolumn{5}{l}{\textbf{Memory Bandwidth Metrics (4)}} \\
BW-001 & Memory Bandwidth Isolation & Bandwidth under contention & \% & Higher \\
BW-002 & Bandwidth Fairness Index & Jain's fairness for bandwidth & 0--1 & Higher \\
BW-003 & Memory Bus Saturation Point & Streams to reach 95\% BW & count & Lower \\
BW-004 & Bandwidth Interference Impact & BW drop from competition & \% & Lower \\
\midrule
\multicolumn{5}{l}{\textbf{Cache Isolation Metrics (4)}} \\
CACHE-001 & L2 Cache Hit Rate & Hit rate under multi-tenant load & \% & Higher \\
CACHE-002 & Cache Eviction Rate & Evictions from other tenants & \% & Lower \\
CACHE-003 & Working Set Collision Impact & Perf drop from cache overlap & \% & Lower \\
CACHE-004 & Cache Contention Overhead & Latency from L2 contention & \% & Lower \\
\midrule
\multicolumn{5}{l}{\textbf{PCIe Bandwidth Metrics (4)}} \\
PCIE-001 & Host-to-Device Bandwidth & H2D transfer rate & GB/s & Higher \\
PCIE-002 & Device-to-Host Bandwidth & D2H transfer rate & GB/s & Higher \\
PCIE-003 & PCIe Contention Impact & BW drop under multi-tenant & \% & Lower \\
PCIE-004 & Pinned Memory Performance & Pinned vs pageable ratio & ratio & Higher \\
\midrule
\multicolumn{5}{l}{\textbf{NCCL/P2P Communication Metrics (4)}} \\
NCCL-001 & AllReduce Latency & Collective allreduce time & $\mu$s & Lower \\
NCCL-002 & AllGather Bandwidth & Allgather achieved bandwidth & GB/s & Higher \\
NCCL-003 & P2P GPU Bandwidth & Direct GPU-to-GPU transfer & GB/s & Higher \\
NCCL-004 & Broadcast Bandwidth & Broadcast collective bandwidth & GB/s & Higher \\
\midrule
\multicolumn{5}{l}{\textbf{Scheduling Metrics (4)}} \\
SCHED-001 & Context Switch Latency & CUDA context switch time & $\mu$s & Lower \\
SCHED-002 & Kernel Launch Overhead & Minimal kernel launch time & $\mu$s & Lower \\
SCHED-003 & Stream Concurrency Efficiency & Concurrent stream efficiency & \% & Higher \\
SCHED-004 & Preemption Latency & High-priority preemption delay & ms & Lower \\
\midrule
\multicolumn{5}{l}{\textbf{Fragmentation Metrics (3)}} \\
FRAG-001 & Fragmentation Index & Memory fragmentation level & \% & Lower \\
FRAG-002 & Allocation Latency Degradation & Latency increase with fragmentation & ratio & Lower \\
FRAG-003 & Memory Compaction Efficiency & Memory reclaimed after defrag & \% & Higher \\
\midrule
\multicolumn{5}{l}{\textbf{Error Recovery Metrics (3)}} \\
ERR-001 & Error Detection Latency & Time to detect CUDA errors & $\mu$s & Lower \\
ERR-002 & Error Recovery Time & Time to recover GPU state & $\mu$s & Lower \\
ERR-003 & Graceful Degradation Score & Resource exhaustion handling & \% & Higher \\
\bottomrule
\end{tabular}
\end{table*}

\section{Usage Examples}

\begin{lstlisting}[language=bash,caption={Common Usage Patterns}]
# Full benchmark on native baseline
./build/gpu-virt-bench --system native --iterations 200

# HAMi-core with specific resource limits
./build/gpu-virt-bench --system hami \
    --memory-limit 2048 \
    --compute-limit 30 \
    --processes 4

# Run only LLM-specific metrics
./build/gpu-virt-bench --system fcsp \
    --metrics LLM-001,LLM-002,LLM-003,LLM-004

# Compare against previous baseline
./build/gpu-virt-bench --system fcsp \
    --compare benchmarks/hami_results.json
\end{lstlisting}


\begin{thebibliography}{99}

\bibitem{brown2020language}
Brown, T., et al. (2020). Language models are few-shot learners. \textit{Advances in Neural Information Processing Systems}, 33.

\bibitem{nvidia_mig}
NVIDIA Corporation. (2024). Multi-Instance GPU User Guide. \url{https://docs.nvidia.com/datacenter/tesla/mig-user-guide/}

\bibitem{hami}
Project HAMi. (2024). Heterogeneous AI Computing Virtualization Middleware. \url{https://github.com/Project-HAMi/HAMi}

\bibitem{vmware_mig}
Sivaraman, H., et al. (2022). Time-sliced vGPU vs MIG vGPU: Choosing the Right vGPU Profile for Your Workload. VMware Technical Report.

\bibitem{jain_fairness}
Jain, R., Chiu, D., \& Hawe, W. (1984). A quantitative measure of fairness and discrimination for resource allocation in shared computer systems. \textit{DEC Research Report TR-301}.

\bibitem{cuda_programming}
NVIDIA Corporation. (2024). CUDA C++ Programming Guide. \url{https://docs.nvidia.com/cuda/cuda-c-programming-guide/}

\bibitem{mlperf}
Mattson, P., et al. (2020). MLPerf: An industry standard benchmark suite for machine learning performance. \textit{IEEE Micro}, 40(2).

\bibitem{transformer}
Vaswani, A., et al. (2017). Attention is all you need. \textit{Advances in Neural Information Processing Systems}, 30.

\bibitem{kubernetes_gpu}
NVIDIA. (2024). Kubernetes GPU Operator. \url{https://github.com/NVIDIA/gpu-operator}

\bibitem{vgpu}
NVIDIA Corporation. (2024). Virtual GPU Software Documentation. \url{https://docs.nvidia.com/grid/}

\end{thebibliography}
\end{document}